\pdfoutput=1
\documentclass[%
aip,  
 amsmath,amssymb,
 reprint,
 floatfix,
]{revtex4-1}
\usepackage[dvipdfmx]{graphicx}
\usepackage{bmpsize}
\usepackage{dcolumn}
\usepackage{bm}

\usepackage[utf8]{inputenc}
\usepackage[T1]{fontenc}
\usepackage{mathptmx}
\usepackage{float}
\usepackage{xcolor}

\usepackage{hyperref}

\DeclareUnicodeCharacter{2009}{\,} 
\begin{document}


\title{An electro-thermal computational study of conducting channels in dielectric thin films using self-consistent phase-field methodology: \\A view toward the physical origins of resistive switching }

\author{Foroozan S. Koushan}%
 \email{foroozan.koushan@gmail.com.}
\author{Nobuhiko P. Kobayashi}%
\affiliation{\leftskip=35pt \rightskip=35pt {Department of Electrical and Computer Engineering, Basking School of Engineering, University of California, Santa Cruz, CA 95064 USA}}%
\affiliation{\leftskip=35pt \rightskip=35pt {Nanostructured Energy Conversion Technology and Research (NECTAR), Univ. of California Santa Cruz, Santa Cruz, CA 95064, USA}}%


\begin{abstract}
\leftskip=35pt \rightskip=35pt {A large number of experimental studies suggest two-terminal resistive switching devices made of a dielectric thin film sandwiched by a pair of electrodes exhibit reversible multi-state switching behaviors; however coherent understanding of physical and chemical origins of their electrical properties needs to be further pursued to improve and customize the performance. In this paper, phase-field methodology is used to study the formation and annihilation of conductive channels resulting in reversible resistive switching behaviors that can generally occur in any dielectric thin films. Our focus is on the dynamical evolution of domains made of electrical charges under the influence of spatially varying electric field and temperature resulting in distinctive changes in electrical conductance.} 
\end{abstract}

\maketitle


Morphological evolution of the interface separating two chemically distinct materials is inherently dynamic, producing distinctive microstructure due to various interactions between interface energy, bulk free-energy, electrothermal, and electrochemical phenomena\cite{n1}. Resistive switching behavior of dielectric thin film is obtained as dynamic evolution of complex microstructures emerging at the interface minimizes the total energy, forming a conductive path resulting in characteristic transport properties of electric charges\cite{n2}. The formation and annihilation of  electrically conductive filaments (ECFs) in dielectric thin films have been studied extensively to understand distinctive functionaries of resistive switching devices that would benefit a variety of technical fields encompassing nonvolatile memories and neuromorphic technology\cite{n3}.

Experimental assessments on ECFs that operate within dielectric thin films have been an important issue in the field of memristive device in recent years. Strachan el al. studied ECFs at the atomic-scale and identified the presence of phase changes that occurred in a functioning TiO$_2$-based memristive device using synchrotron-based x-ray absorption spectromicroscopy and transmission electron microscopy\cite{n4}, providing experimental evidence of regions of dynamic conductivity modulations during resistive switching. In another study, Nallagatla el al. used an Au-coated probe tip as a mobile top electrode to show that localizing electric field and using an epitaxial homogeneous thin-film improved the performance and repeat-ability of resistive switching\cite{n5}. 

Along with these experimental assessments, the nature of ECFs has been investigated extensively using various computational approaches that employ phenomenological electro-thermal modeling often in conjunction with the finite-element method\cite{n10,n11}. In many existing models, the formation and annihilation of ECFs are resolved on the basis of solutions obtained self-consistently for the continuum transport equations that dynamically emulate advection and diffusion of electrical charges under the influence of electric field and temperature. In other words, the presence – rather than the formation – of a single ECF is set as the initial condition, and then, the dependence of physical properties (i.e., electrical conductance) of the pre-conditioned ECF on such parameters as electrical potential and temperature is evaluated to reproduce unique electrical characteristics of resistive switching. As a result, these models heavily rely on diffusion and mobility formulations based on the transport of, for instance, the Frenkel-type vacancies\cite{n6}. For instance, Strukove\cite{n7} first proposed a linear and then later an exponential ion drift mobility, and Yang\cite{n8} suggested a non-linear dependence of dopant drifts under applied voltage. A ECF combined with a gap was also used along with stochastic properties of ions having spatial variations within the ECF\cite{n9}. Furthermore, an electro-thermal modeling was carried out for a device represented by a cylindrical shape that contained a pre-existing single ECF assumed to have a plain cylindrical shape\cite{n10}. When various thermal processes are included in the formulation of ECFs, the Fourier equation is used to model annihilation of ECFs by invoking Joule heating, which was found to be consistent with the temperature of ECFs, experimentally obtained\cite{n14}. In multiple studies, an electro-thermal model in conjunction with the finite-element method was performed by using the migration of such electrical charges as oxygen vacancies\cite{n1,n2,n3,n11}. A similar method was used for a structure that consists of heterogeneous bi-layered dielectric stack\cite{n12}. Moreover, multiple ECFs were considered to participate in electrical properties to explain multi-level switching characteristics\cite{n12,n13}. 

In this paper, the phase-field method\cite{n15,n16} is employed to explicitly illustrate the formation and annihilation of ECFs in a dielectric thin film to assess characteristic electrical properties of resistive switching devices. The previous studies described earlier all assumed the presence of a single ECF that has specific dimensions and geometrical morphologies in conjunction with general descriptions on the transport of electrical charges under the influence of electric potential and heat. In contrast, our approach based on the phase-field method tracks the dynamical formation and evolution of domains (i.e., ECFs) made of electrical charges present in a dielectric thin film as the energy associated with the system – a dielectric thin film containing electrical charges forming domains – is reduced over time. While the phase-field method successfully predicted the formation of multiple ECFs under the influence of electric field externally applied under isothermal conditions in our previous studies\cite{n17}, this paper presents a substantial extension of our previous study by including the dependence of the formation and annihilation of ECFs on local electrical potential and local temperature to elucidate the on and off states of a resistive switching device. 

\begin{figure}
    \centering
    \includegraphics[width=0.45\textwidth]{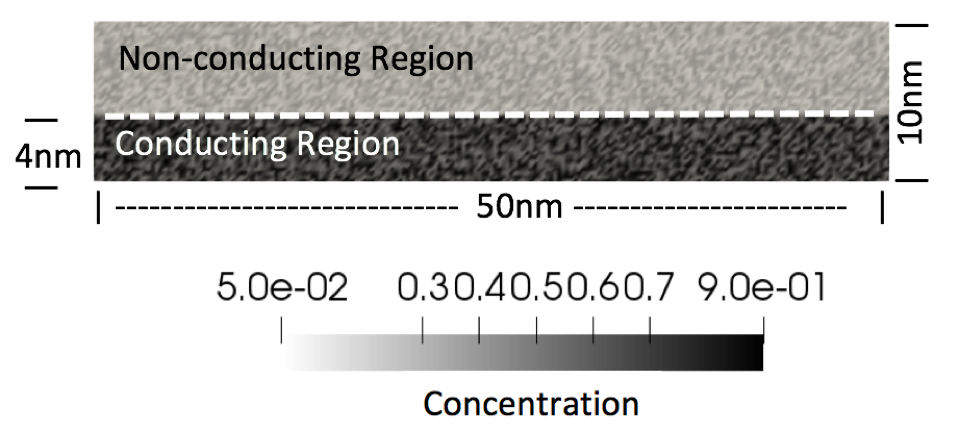}
    \caption{A 50 nm x 10 nm dielectric thin film composed of a conducting region with initial concentration in the range of 0.7-0.9 and a non-conduction region with concentration in the range of 0.1-0.3.}
    \label{fig:one}
\end{figure}

Molecular dynamics applied to the current context would track the motion of charged particles present in a dielectric thin film in a resistive switching device, in contrast, the phase-field method used in our study tracks the dynamical evolution of clusters made of charged particles in a dielectric thin film. These charge-clusters that represent regions of electrically conducting collectively define geometrical boundaries (i.e., interfaces) that separate the electrical conducting regions from a non-conducting region, allowing us to avoid the necessity of expressing dynamic boundary conditions over an evolving interface and significantly reducing computational burdens without forfeiting overall integrity of the modeling.
In our study, a 50 nm x 10 nm dielectric thin film illustrated in Fig.~\ref{fig:one} is considered as a system. This system is viewed as an as-fabricated resistive switch comprising a dielectric thin film in which two distinct regions separated by an interface, exemplifying a resistive switch made of a dielectric thin film in which mobile charges are initially distributed in a certain way by which the two regions, one conducting and the other non-conducting are formed. As illustrated in Fig.~\ref{fig:one}, the system consists of two distinctive regions: a conducting region with normalized concentration of charged particles $c_({\bf r}, t)$, where ${\bf r}$ is a position vector within the system and $t$ is time, set to a random local value varied in the range of 0.7-0.9 at $t$ = 0, and a non-conducting region with  $c_({\bf r}, t)$ set to a random local value chosen in the range of 0.1-0.3 at $t$ = 0, separated by an interface along which $c$ also varies randomly. In our previous study\cite{n17}, the presence of the non-uniformity in $c$ along the interface between the two regions was found to induce the formation of ECFs under the influence of external electric potential . This specific initial structure allows a double-well free-energy density function and the diffuse interface approximation to suitably describe dynamical structural evolution of the dielectric thin film. The bulk free-energy density function $f_{bulk}(c_{({\bf r},t)})$ associated with the system in 
Fig.~\ref{fig:one} is given by:
\begin{eqnarray}
 f_{bulk} (c_{({\bf r},t)})=&&{A}{\bf [}c_{({\bf r},t)}-c_1{\bf ]}^2{\bf [}c_{({\bf r},t)}-c_2{\bf ]}^2 
\label{eq:one}
\end{eqnarray}
where $A$ is the magnitude of the double-well potential, $c_{1}$ and $c_{2}$ are the normalized concentrations of conducting and non-conducting states. Variable $c_{({\bf r},t)}$ in Eq.~(\ref{eq:one}) is bounded to $0{\bf <} c_{({\bf r},t)}{\bf <}1$, where zero corresponds to pure non-conducting state and unity corresponds to pure conducting state. 

In a dielectric thin film structure, with specific dimensions set in our modeling, the bulk free energy density of the system is expected to be dominated by the total surface tension experienced by the entire conductive regions made of charged particles. In this view, the dependence of the surface tension on temperature $T$ is translated into the dependence of the bulk free energy density on temperature in the modeling. The dependence of the surface tension on temperature shown by Guggenheim-Katayama\cite{n19} is expressed as follows:  
\begin{eqnarray}
 \gamma=&&\gamma^o{\bf (}1-\frac{T}{T_c}{\bf )}^n 
\label{eq:two}
\end{eqnarray}
where ${\gamma^o}$ is the concentration-dependent surface tension, ${n}$ is an empirical factor, and ${T_c}$ is the critical temperature of the system at which the phase of the material changes. So as temperature $T$ raises above ${T_c}$, the thin-film undergoes an irreversible phase change impacting its electrical characteristics in such a way that the system no longer works as a resistive switching device. Eq.~(\ref{eq:two}) suggests that surface tension, and thus its corresponding surface energy, decreases as the temperature ${T}$ increases, and becomes zero when ${T}$ matches ${T_c}$. Empirical factor ${n}$ that depends on a specific material system modulates the rate at which ${\gamma}$ responds to changes in ${T}$. In other words, ${n}$ determines the order of the polynomial equation relating ${\gamma}$ to ${T}$, and its specific value is obtained experimentally. In this paper, ${n}$ is set to 2 as it usually takes a value in the range between 1 and 2 for liquids\cite{n25}.

Although Eq.~(\ref{eq:two}) conventionally applies to an interface that separates two liquid phases, in this paper we are assuming that utilizing the dependence of ${\gamma}$ on ${T}$ as in Eq.~(\ref{eq:two}) is deemed valid for interfaces of two solid phases – conducting and non-conducting regions as shown in Fig.~\ref{fig:one} – present in a dielectric thin film, because of virtual melting theorized for solid-solid interfaces at the nano-meter scale\cite{n27}. Therefore, in our modeling, free energy density of the system in Fig.~\ref{fig:one} is considered to be equivalent to the total surface tension present within the system. This treatment allows us to incorporate the T-dependence provided in Eq.~(\ref{eq:two}) into Eq.~(\ref{eq:one}), resulting in an expression for temperature-dependent bulk free-energy density function: 
\begin{eqnarray}
 f_{\text{bulk}}(c_{({\bf r},t)},T)=&&{A}{\bf [}c_{({\bf r},t)}-c_1{\bf ]}^2{\bf [}c_{({\bf r},t)}-c_2{\bf ]}^2 {\bf (}1-\frac{T}{T_c}{\bf )}^n 
\label{eq:three}
\end{eqnarray}

In general, the phase-field method is a natural extension of diffuse-interface models by Cahn and Allen\cite{n20}, Ginzburg and Landau\cite{n21}, and Cahn and Hilliard\cite{n22}. The equations involved in the phase-field approach are developed in terms of the following free-energy function:
\begin{eqnarray}
 F=&&\int_{R}[f_{\text{bulk}}(c_{({\bf r},t)},T)+\frac{1}{2}\epsilon |  \nabla c_{({\bf r},t)}|^2]dr
\label{eq:four}
\end{eqnarray}
\\where $R$ is the entire area of the system, $f_{bulk}(c_{({\bf r},t)}, T)$ is the free-energy density function given in Eq.~(\ref{eq:three}), $c_{({\bf r},t)}$ is the concentration of charged particles. 
Also in Eq.~(\ref{eq:four}), $\epsilon$ is the interfacial gradient energy term relating to the energy stored per unit application of the potential of the gradient of  $c_{({\bf r},t)}$. This term represents potential energy of the interface, and it is assumed to be uniformly constant along the interface.

The dynamical evolution of charge-clusters as the system establishes thermal equilibrium is obtained by invoking a general form of phase-field conservation law, as the first-order variation of free-energy functional equation results in phase-field flux:
\begin{eqnarray}
 \frac{\partial c_{({\bf r},t)}}{\partial t}=&& \nabla . (M  \nabla \frac{F}{c_{({\bf r},t)}})
\label{eq:five}
\end{eqnarray}
where $M$ is the mobility of the conserved variable representing a physical property of the system, and is assumed to be constant.
Eq.~(\ref{eq:five}) is a Cahn-Hilliard phase field equation in concentration variable $c_{({\bf r},t)}$ for the system, which incorporates effect of temperature on the movement of charge-clusters within the system. Self-consistent solutions to this equation are obtained by using the Multi-physics Object-Oriented Simulation Environment (MOOSE) finite-element platform \cite{n23,n24}. 

\begin{figure}
    \scalebox{.5}{\includegraphics{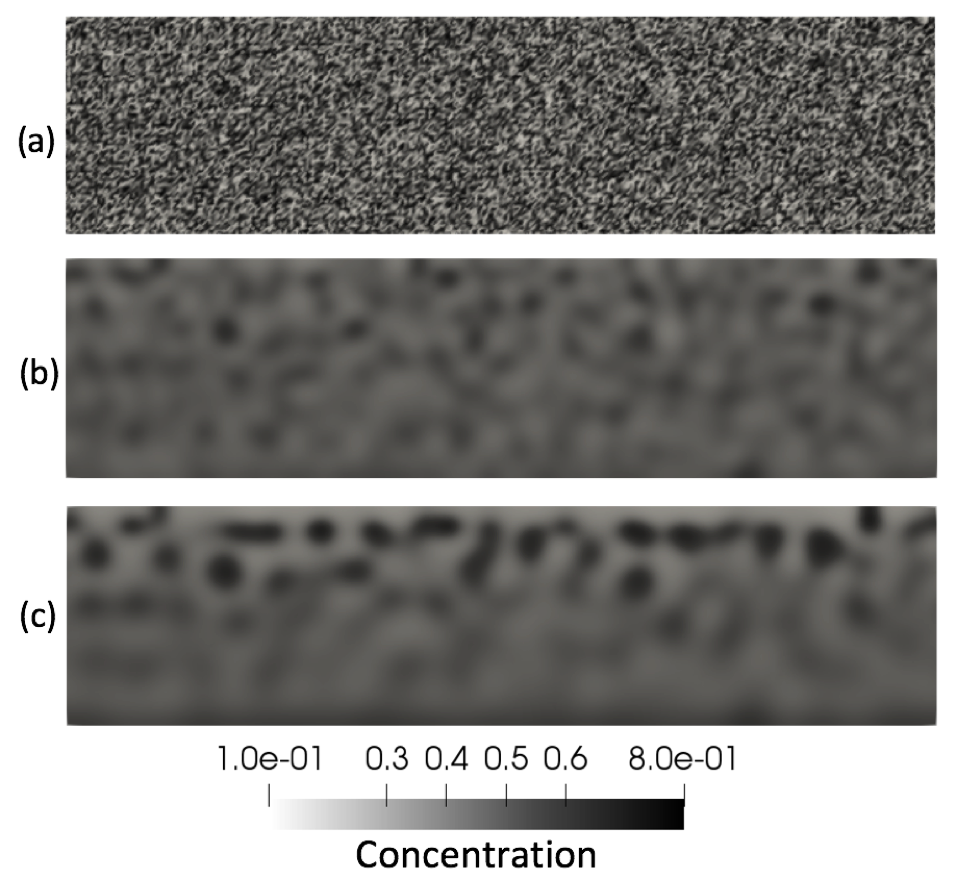}}
    \caption{ Effect of external temperature gradient on Spinodal decomposition of a system with 50 nm x 40 nm dimension, consisting of conduction and non-conducting materials. Panel (a) represents the initial state at $t_0$. A temperature gradient was applied in such a way that bottom was at higher temperature compared to that of the top. Evolution of charge-clusters was captured at different times, $t$=38 s and $t$=144 s, in panel (b) and (c), respectively.}
    \label{fig:two}
\end{figure}

To elucidate the sole effect of temperature $T$ on the evolution of charge-clusters, a system of an 80 nm x 40 nm dielectric thin film, as shown in Fig.~\ref{fig:two}, was considered. The initial charge concentration $c_({\bf r},t)$ at $t$ = 0 was varied in the range between 0.1 and 0.8 randomly across the system. 
Dimensions for this system were set to values much larger than those of the system drawn in Fig.~\ref{fig:one} so that variations in the grey-scale across the figure appears visible and distinct. A temperature gradient difference of 50 K was applied between the top and the bottom surface of the system; with top being set at temperature lower than that of the bottom by 50 K. Periodic boundary conditions were imposed on the left and the right side of the system for $c_{({\bf r},t)}$ and Dirichlet boundary conditions were applied to the top and the bottom of the system for $T$. Simulation time was kept constant for all the panels in Fig.~\ref{fig:two} by purposely halting the computation before the system reached the absolute steady-stated at which all charged particles aggregated near the top of the system. The dynamic evolution of the clusters was captured at different time intervals as shown in panels (b) and (c), highlighting the effect of temperature to the evolution – in shape and size – of charge-clusters at a given simulation time. These figures clearly show how a temperature gradient, with absence of electrical potential, contributes to defining the cluster boundaries as total free energy of the system reduces. High temperature forms clusters with less defined boundaries, whereas the boundary and shape of clusters formed at lower temperature are more well distinct. In the next section, electrical potential is added to the temperature gradient. 

After evaluating the effect of $c_({\bf r},t)$ on temperature, the effects of applying external electric potential was assessed in an isothermal environment. The dependence of bulk free energy, $F$, on the external electric potential $V_({\bf r},t)$ is incorporated by adding electrostatic energy $g_{elec}$ to Eq.(\ref{eq:four}), resulting in the following formulation, where the external electric potential is coupled to the electric Laplace equation:
\begin{eqnarray}
 F=&&\int_{R}[f_{\text{bulk}}(c_{({\bf r},t)},T)+\frac{1}{2}\epsilon |  \nabla c_{({\bf r},t)}|^2\nonumber\\ 
 &&+ g_{elec}(c_{({\bf r},t)},V)]dr   
\label{eq:six}
\end{eqnarray}
$g_{elec}$ is defined as,
\begin{eqnarray}
g_{elec}(c_{({\bf r},t)},V)=&&\frac{q}{\Omega}V_{({\bf r},t)}c_{({\bf r},t)}
\label{eq:seven}
\end{eqnarray}
where $V_{({\bf r},t)}$ is the electrical potential applied across a dielectric thin film, $q$ is the electric charge, and $\Omega$ is the differential volume unit of the mesh cell used in the system.
Self-consistent solutions for $c_{({\bf r},t)}$ can be calculated by incorporating Eq.~(\ref{eq:six}) and Eq.~(\ref{eq:seven}) into Eq.~(\ref{eq:five}),self-consistently coupled to electronic Laplace equation\cite{n26,n17}: 
\begin{eqnarray}
    \nabla.\sigma(c_{({\bf r},t)}) \nabla V_{({\bf r},t)} =&&0
    \label{eq:eight}
\end{eqnarray}
where electrical conductivity $\sigma$ is a linear function of variable $c_{(\bf r,t)}$. 
\begin{figure}
    \scalebox{.48}{\includegraphics{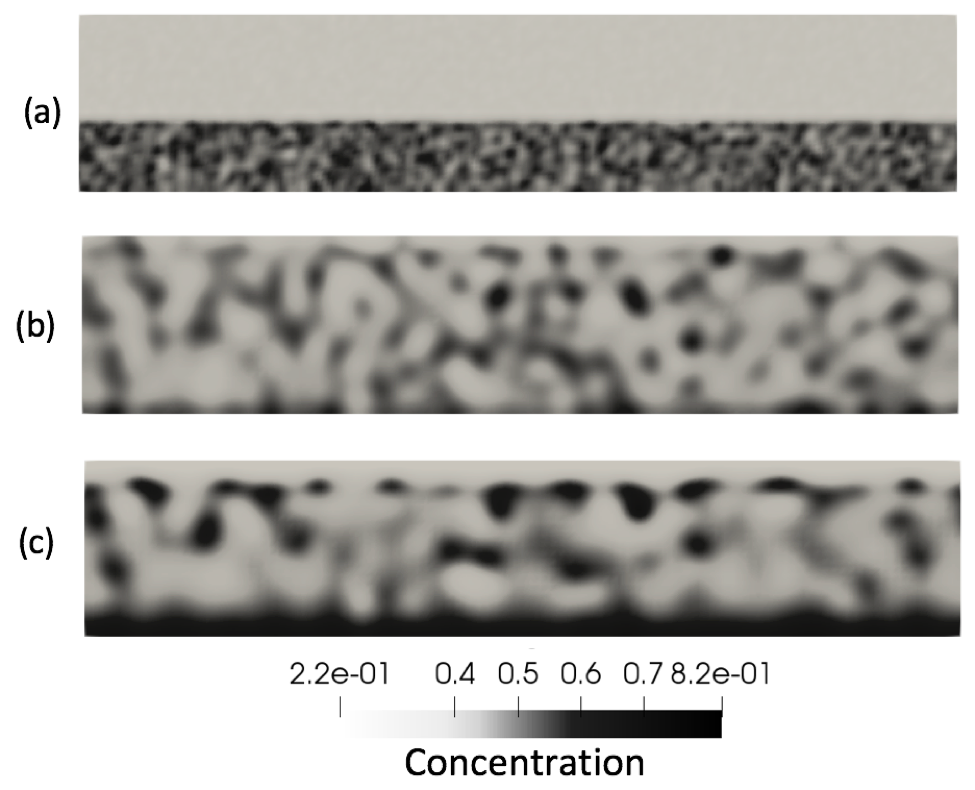}}
    \caption{The system consisting of dielectric thin film with dimension of 50 nm x 10 nm at 400 K. (a) pristine-state, (b) programmed-state (ON-state) established by applying electrical potential and (c) erased-state (OFF-state) obtained subsequently by reversing the electrical potential.}
    \label{fig:three}
\end{figure}
Solving the phase field conservation law allows us to demonstrate the formation and annihilation of ECFs in a dielectric thin film as the system seeks the minimum bulk free energy under the influence of the external electrical potential at a specific temperature and thermal force. The system used for the electro-thermal modeling is illustrated in Fig.~\ref{fig:three}(a) that represents a dielectric thin film at its pristine state. In Fig.~\ref{fig:three}(a), a conductive region - the upper region with uniform light contrast - with high $c_({\bf r},t)$ is separated from a non-conductive region - the lower region with non-uniform contrast - with low $c_({\bf r},t)$ by an interface along which $c_({\bf r},t)$ varies. Electrical resistance, along the thickness of the system in the pristine-state, is the highest throughout the lifetime of the system. Fig.~\ref{fig:three}(b) represents the programmed-state established when $V$=1V was applied to the top electrode while the bottom electrode was set to 0V, and the temperature was set to 400 $K$. Under these conditions, multiple ECFs were formed, connecting the top and the bottom electrodes, resulting in low electrical resistance (i.e., ON-state). Fig.~\ref{fig:three}(c) represents the erased-state, and it was achieved when the polarity of the external electric potential was reversed. Fig.~\ref{fig:three}(c) shows that the ECFs were annihilated, resulting in high electrical resistance (i.e., OFF-state). The fact that the thickness of the non-conductive region near the top electrode in Fig.~\ref{fig:three}(c) is smaller than that of the non-conductive region in Fig.~\ref{fig:three}(a) indicates that electrical resistance of the OFF-state in Fig.~\ref{fig:three}(c) is lower than that in the pristine-state in Fig.~\ref{fig:three}(a),which is consistent with RRAM and memristor devices that experimentally exhibit initial electrical resistance at the pristine-state much higher than that of the OFF-state. 
\\
\begin{figure}
    \scalebox{.69}{\includegraphics{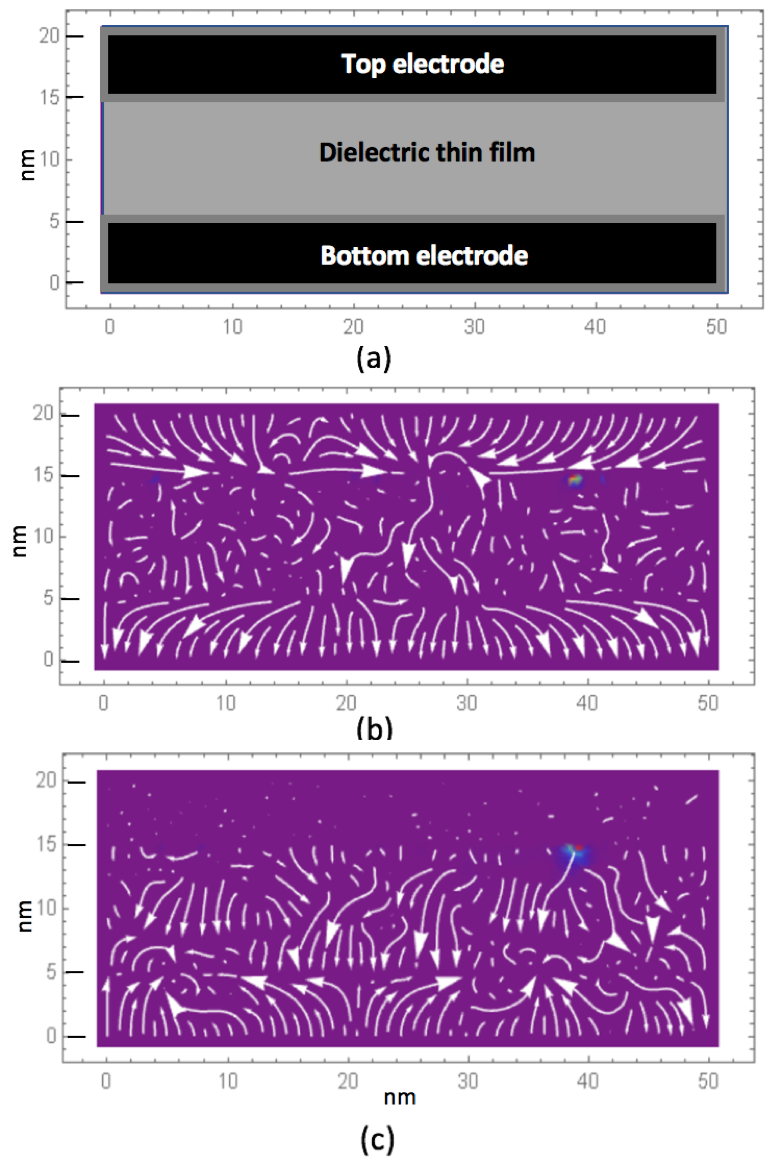}}
    \caption{Electrical current density maps for ON- and OFF-state were obtained by building the system shown in (a). In (b) and (c), local electrical current density is represented by an arrow with length proportional to the relative magnitude of local electrical current density in (b) ON-state, and (c) OFF-state.}
    \label{fig:four}
\end{figure}
The $c_{({\bf r},t)}$ maps in Fig.~\ref{fig:three}(b) and (c) are further converted into maps of electrical current density to portray how electrical current maneuvers in the two different states, ON-state and OFF-state, and to elucidate how the formation and annihilation of multiple ECFs with complex shapes are reflected in the flow of electrical current. Fig.~\ref{fig:four}(a) illustrates how the system used for mapping electrical current density was built. The system consists of a dielectric thin film sandwiched by a pair of 5 nm thick metal electrodes. the dielectric thin film was replaced by either the one in the ON-state shown in Fig.~\ref{fig:four}(b) or the other in the OFF-state presented in Fig.~\ref{fig:four}(c), resulting in current density maps for the ON-state in Fig.~\ref{fig:four}(b) and for the OFF-state in Fig.~\ref{fig:four}(c). The electrical current density maps were obtained by applying $V$ =100 mV to the top electrode while the bottom electrode was set to 0 V.
When the system is in the ON-state in Fig.~\ref{fig:four}(b), a long arrow connecting top electrode to the bottom electrode clearly indicate the presence of a dominant conductive path that funnels electrical current flowing laterally across the top electrode. In contrast, as the electric potential was reserved, and thus, the system entered the OFF-state, all the arrows present in the top electrode in Fig.~\ref{fig:four}(b) disappeared as seen in Fig.~\ref{fig:four}(c), indicating that the flow of electrical current discontinued. 
\begin{figure}
    \centering
    \includegraphics[width=0.44\textwidth]{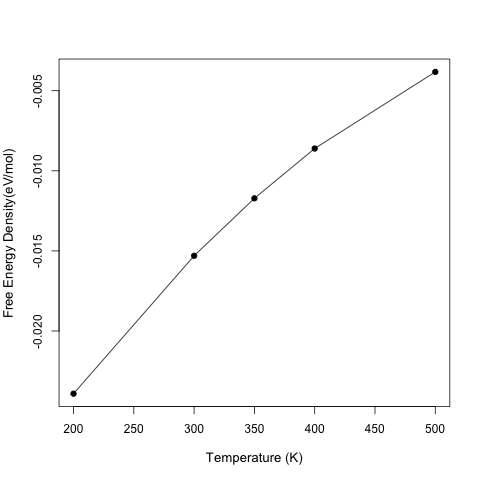}
    \caption{ The free energy density of the system at various temperatures. In this simulation, critical Temperature $T_c$ was set to 700K, and empirical factor $n$ was set to 2 in Eq.(\ref{eq:two}). The number of iterations to reach the convergence criteria for a self-consistence solution to Eq.(\ref{eq:six}) and (\ref{eq:seven}) was fixed so that the system did not reach the ground state at various temperatures.}
    \label{fig:five}
\end{figure}
Fig.~\ref{fig:three} and Fig.~\ref{fig:four} were obtained for $T$=400K. 

In order to evaluate the overall endurance of resistive switching, which is expected to depend on the stability of the formation and annihilation of ECFs, the effect of increasing temperature on the free energy density of the system was assessed. In the assessment, the number of computational iterations was fixed in reducing the free energy density for the pristine-state shown in Fig.~\ref{fig:three}(a) at various temperatures. In other words, at various temperatures, the iteration process was stopped when a certain number of iterations was completed and a reduction in the free energy density was recorded. Fig.~\ref{fig:five} shows reductions in the free energy obtained at various temperatures, indicating that reduction in the free energy density gradually decreases as the temperature increases, that is to say that as the temperature increases, it takes longer time for the system to reach minimum free energy. Thus, the stability of the system – reversible and repeatable formation and annihilation of ECFs – reduces as the temperature rises. This trend is valid as long as the temperature is well below $T_{c}$ of the system. In other words, as the $T_{c}$ is expected to be determined by a range of factors including material properties, it needs to be as high as possible for the device to achieve high stability at a larger temperature range.

In summary, electro-thermal characteristics of the formation and annihilation of ECFs in a system consisting of dielectric thin film were studied by using the phase-field method. The dependence of interface energy on local temperature and variations in local concentration of electrical charges were formulated and incorporated in describing bulk free energy density of the system in our treatment based on the Cahn-Hilliard model. Our results suggest that as the system temperature increases, the formation of charge-clusters with weakly defined boundaries is promoted, which also translates into that the free energy density of the system reduces at slower rates as the system temperature drifts higher. In other words, the stability of the system - reversible and repeatable formation and annihilation of ECFs - reduces as the temperature rises, extending significant implications for endurance of resistive switching devices and memristors. Our results clearly illustrate that the formation and annihilation of ECFs are distinctly accomplished by controlling electrical potential applied across the system as resistive switches and memristors normally operate. The resulting electrical current density maps provide a realistic representation of a complex conducting path through which electrical current flows in the ON-state and a cessation of the electrical current when ECFs rupture in the OFF-state as the polarity of applied electrical potential is reversed across the system. A significant difference in the thickness of non-conductive regions in the pristine-state and in the OFF-state highlights that the irreversible characteristics of a process often referred to as electroforming result in the resistance of the OFF-state to be always smaller than that of the pristine-state. 
\\

The data that support the findings of this study are available from the corresponding author, upon reasonable request.

\bibliography{ET-paper}

\end{document}